\documentclass[ sort&compress,final]{aipproc}

\layoutstyle{8x11double}

\begin{document}
%
%  These Macros are taken from the AAS TeX macro package version 4.0.
%  Include this file in your LaTeX source only if you are not using
%  the AAS TeX macro package and need to resolve the macro definitions
%  in the BibTeX entries returned by the ADS abstract service.
%
%  For more information on the AASTeX macro package, please see the URL
%	http://www.aas.org/publications/aastex.html
%  For more information about ADS abstract server, please see the URL
%	http://adswww.harvard.edu/ads_abstracts.html
%

% Abbreviations for journals.  The object here is to provide authors
% with convenient shorthands for the most "popular" (often-cited)
% journals; the author can use these markup tags without being concerned
% about the exact form of the journal abbreviation, or its formatting.
% It is up to the keeper of the macros to make sure the macros expand
% to the proper text.  If macro package writers agree to all use the
% same TeX command name, authors only have to remember one thing, and
% the style file will take care of editorial preferences.  This also
% applies when a single journal decides to revamp its abbreviating
% scheme, as happened with the ApJ (Abt 1991).

\def\jnl@style{\it}
%commente par Seb
\def\aaref@jnl#1{{\jnl@style#1}}
%ref remplace par aaref pour eviter conflit...

\def\aaref@jnl#1{{\jnl@style#1}}

\def\aj{\aaref@jnl{AJ}}                   % Astronomical Journal
\def\araa{\aaref@jnl{ARA\&A}}             % Annual Review of Astron and Astrophys
\def\apj{\aaref@jnl{ApJ}}                 % Astrophysical Journal
\def\apjl{\aaref@jnl{ApJ}}                % Astrophysical Journal, Letters
\def\apjs{\aaref@jnl{ApJS}}               % Astrophysical Journal, Supplement
\def\ao{\aaref@jnl{Appl.~Opt.}}           % Applied Optics
\def\apss{\aaref@jnl{Ap\&SS}}             % Astrophysics and Space Science
\def\aap{\aaref@jnl{A\&A}}                % Astronomy and Astrophysics
\def\aapr{\aaref@jnl{A\&A~Rev.}}          % Astronomy and Astrophysics Reviews
\def\aaps{\aaref@jnl{A\&AS}}              % Astronomy and Astrophysics, Supplement
\def\azh{\aaref@jnl{AZh}}                 % Astronomicheskii Zhurnal
\def\baas{\aaref@jnl{BAAS}}               % Bulletin of the AAS
\def\jrasc{\aaref@jnl{JRASC}}             % Journal of the RAS of Canada
\def\memras{\aaref@jnl{MmRAS}}            % Memoirs of the RAS
\def\mnras{\aaref@jnl{MNRAS}}             % Monthly Notices of the RAS
\def\pra{\aaref@jnl{Phys.~Rev.~A}}        % Physical Review A: General Physics
\def\prb{\aaref@jnl{Phys.~Rev.~B}}        % Physical Review B: Solid State
\def\prc{\aaref@jnl{Phys.~Rev.~C}}        % Physical Review C
\def\prd{\aaref@jnl{Phys.~Rev.~D}}        % Physical Review D
\def\pre{\aaref@jnl{Phys.~Rev.~E}}        % Physical Review E
\def\prl{\aaref@jnl{Phys.~Rev.~Lett.}}    % Physical Review Letters
\def\pasp{\aaref@jnl{PASP}}               % Publications of the ASP
\def\pasj{\aaref@jnl{PASJ}}               % Publications of the ASJ
\def\qjras{\aaref@jnl{QJRAS}}             % Quarterly Journal of the RAS
\def\skytel{\aaref@jnl{S\&T}}             % Sky and Telescope
\def\solphys{\aaref@jnl{Sol.~Phys.}}      % Solar Physics
\def\sovast{\aaref@jnl{Soviet~Ast.}}      % Soviet Astronomy
\def\ssr{\aaref@jnl{Space~Sci.~Rev.}}     % Space Science Reviews
\def\zap{\aaref@jnl{ZAp}}                 % Zeitschrift fuer Astrophysik
\def\nat{\aaref@jnl{Nature}}              % Nature
\def\iaucirc{\aaref@jnl{IAU~Circ.}}       % IAU Cirulars
\def\aplett{\aaref@jnl{Astrophys.~Lett.}} % Astrophysics Letters
\def\apspr{\aaref@jnl{Astrophys.~Space~Phys.~Res.}}
                % Astrophysics Space Physics Research
\def\bain{\aaref@jnl{Bull.~Astron.~Inst.~Netherlands}} 
                % Bulletin Astronomical Institute of the Netherlands
\def\fcp{\aaref@jnl{Fund.~Cosmic~Phys.}}  % Fundamental Cosmic Physics
\def\gca{\aaref@jnl{Geochim.~Cosmochim.~Acta}}   % Geochimica Cosmochimica Acta
\def\grl{\aaref@jnl{Geophys.~Res.~Lett.}} % Geophysics Research Letters
\def\jcp{\aaref@jnl{J.~Chem.~Phys.}}      % Journal of Chemical Physics
\def\jgr{\aaref@jnl{J.~Geophys.~Res.}}    % Journal of Geophysics Research
\def\jqsrt{\aaref@jnl{J.~Quant.~Spec.~Radiat.~Transf.}}
                % Journal of Quantitiative Spectroscopy and Radiative Transfer
\def\memsai{\aaref@jnl{Mem.~Soc.~Astron.~Italiana}}
                % Mem. Societa Astronomica Italiana
\def\nphysa{\aaref@jnl{Nucl.~Phys.~A}}   % Nuclear Physics A
\def\physrep{\aaref@jnl{Phys.~Rep.}}   % Physics Reports
\def\physscr{\aaref@jnl{Phys.~Scr}}   % Physica Scripta
\def\planss{\aaref@jnl{Planet.~Space~Sci.}}   % Planetary Space Science
\def\procspie{\aaref@jnl{Proc.~SPIE}}   % Proceedings of the SPIE

\let\astap=\aap
\let\apjlett=\apjl
\let\apjsupp=\apjs
\let\applopt=\ao

\title{Exploring the origin of neutron star magnetic field: magnetic properties of the progenitor OB stars}


\author{V\'eronique Petit}{address={D\'ept. de Physique, Universit\'e Laval, Qu\'ebec (QC), Canada, G1K 7P4},altaddress={Observatoire du Mont-M\'egantic, Qu\'ebec, Canada} }

\author{Gregg A. Wade}{address={Dept. of Physics, Royal Military College of Canada, PO Box 17000, Stn Forces, Kingston, Canada, K7K 4B4}}

\author{Laurent Drissen}{address={D\'ept. de Physique, Universit\'e Laval, Qu\'ebec (QC), Canada, G1K 7P4},altaddress={Observatoire du Mont-M\'egantic, Qu\'ebec, Canada}}

\author{Thierry Montmerle}{address={Laboratoire d'Astrophysique de Grenoble, Universit\'e Joseph-Fourier, 38041 Grenoble Cedex 9, France}}

\begin{abstract}
 Ferrario \& Wickramasinghe (2006) explored the hypothesis that the magnetic fields of neutron stars are of fossil origin. In this context, they predicted the field distribution of the progenitor OB stars, finding that 5 per cent of main sequence massive stars should have fields in excess of 1\,kG. We have carried out sensitive ESPaDOnS spectropolarimetric observations to search for direct evidence of such fields in all massive B- and O-type stars in the Orion Nebula Cluster star-forming region.
We have detected unambiguous Stokes V Zeeman signatures in spectra of three out of the eight stars observed (38\%). Using a new state-of-the-art Bayesian analysis, we infer the presence of strong (kG), organised magnetic fields in their photospheres. For the remaining five stars, we constrain any dipolar fields in the photosphere to be weaker than about 200\,G.  Statistically, the chance of finding three $\sim1\rm\,kG$ fields in a sample of eight OB stars is quite low (less than 1\%) if the predictions of Ferrario \& Wickramasinghe are correct. This implies that either the magnetic fields of neutron stars are not of fossil origin, that the flux-evolution model of Ferrario \& Wickramasinghe is incomplete, or that the ONC has unusual magnetic properties. We are undertaking a study of other young star clusters, in order to better explore these possibilities.
\end{abstract}

\maketitle

\section{Introduction}

The origin of the magnetic fields of massive stars, present at all evolutionary stages (PMS, MS, post-MS \cite{Alecian2007MNRAS,Borra1980ApJS42p421,Schmidt2003ApJ595p1101}), remains an open question. 
Two possible models may explain the presence of magnetic fields: 

(i) In the dynamo model, the field is generated by a dynamo mechanism, occurring classically in convective regions or induced by strong shear during differential rotation. 

(ii) In the fossil model, the field is a remnant from a dynamo active during a previous evolutionary phase, or swept up from the interstellar medium (ISM) during star formation. This scenario implies that the field must somehow survive the various structural changes encountered during stellar evolution. The magnetic flux is usually assumed to be conserved to some extent.

The dynamo model reproduces well the characteristics of late-type main sequence stars and giants. 
However, it fails to explain the fields of magnetic Ap stars,  as their envelopes are entirely radiative.
Some models of convection in the small convective core of those stars have been put forward, but they still have fundamental difficuties reproducing the observed fields \citep{Charbonneau2001ApJ559p1094}.
Their simple magnetic geometries, lack of significant mass-field strength or period-field strength relation, and the fact that the observed characteristics of Herbig star magnetic fields \citep[e.g.][]{Wade2007MNRAS376p1145,Wade2005AA442pL31,Catala2007AA462p293,Alecian2006ArXiv0612186,Folsom2007MNRAS376p361} are qualitatively identical to those of Ap stars, point toward a fossil origin.

Furthermore, the incidence, geometries and strengths of white dwarf magnetic fields are at least qualitatively compatible with evolution from Ap-Bp stars, suggesting that the fields of white dwarfs may also be of fossil origin \citep{Wickramasinghe2005MNRAS356p1576}. 

Similar questions can be asked about neutron stars. Neutron star field strengths, inferred from spin down rates of radio pulsars, are in the range of $10^{11}$-$10^{14}\rm\,G$. A particular class of pulsars called the magnetars, which include the anomalous X-ray pulsars and soft gamma repeaters, have fields that range up to $10^{14}$-$10^{15}\rm\,G$.
The two possible scenarios that have been put forward to explain the existence of these fields are again the fossil and dynamo mechanisms. 

The fossil hypothesis implies that neutron star fields come from the progenitor OB star fields which have survived the post main sequence and the core collapse phases. 
There is some observational evidence that neutron stars may evolve from stars as massive as $45\rm\,M_\odot$ \citep{Gaensler2005ApJ620pL95,Muno2006ApJ636pL41}. 
The only two known O stars with directly measured magnetic fields are $\theta^1\rm\,Ori\,C$ and HD\,191612 ($\sim40\rm\,M_\odot$), with dipolar field strengths ranging between 1000-1500\,G \citep{Donati2002MNRAS333p55,Donati2006MNRAS365pL6}. 
The magnetic flux of these stars ($\sim10^{27}\rm\,G\,cm^2$) is of the same scale as the magnetic flux of the highest field magnetars. 

Provided that OB star fields are remnants from the ISM, the fossil hypothesis could provide a powerful explanation of the wide range of magnetic fields present in neutron stars, and may also explain the super strong fields seen in magnetars.

In this fossil model, the properties of neutron star magnetic fields are a function of the field properties of the progenitor OB stars, plus any physics that occurs during post main sequence evolution to alter those fields. Depending on the relative importance of these two ingredients, the fields of neutron stars might be nearly identical to those of OB stars (but stronger, of course), or rather different. 
 
On the other hand, it has been suggested that neutron star magnetic fields could instead be generated by a dynamo mechanism taking place during the core collapse itself, and induced by differential rotation. 
Present studies \citep{Heger2005ApJ626p350} assume that any primordial fields present in the progenitor star are weak enough to be expelled by the dynamo process. 
However, if the initial field is strong enough, the evolution will be different, as this field is likely to interfere with the differential rotation and therefore with the dynamo process itself \citep{Spruit1999AA349p189}.

Hence, in both cases, and the likely combination of them, a primordial field present during the formation of a neutron star will play a fundamental role. Knowing the magnetic properties of the progenitor OB stars would therefore be an important asset for constraining models of stellar evolution leading to neutron star birth.
 
\section{Observations}

Magnetic fields can be directly detected in stellar atmospheres by the means of the Zeeman effect. 
If the field is strong enough, and the spectral lines narrow enough, one can directly see the Zeeman splitting of the lines in the intensity spectrum.
However, if the field is weaker, and the lines broadened either intrinsically or by fast stellar rotation, the splitting is much more difficult to detect, even at high spectral resolution. 

In that case, the most effective way to detect the Zeeman effect is by looking at circular polarisation signatures across photospheric spectral lines. Recently, a technique called "Least Squares Deconvolution" (LSD) has been developed by \citet{Donati1997MNRAS291p658} for extracting a mean Stokes V profile, using all the lines present in a spectrum simultaneously. This allows for better detection limits as it substantially increases the signal-to-noise ratio. This technique has been proven to be very useful for detecting magnetic fields in Ap-Bp stars \citep{Auriere2007} and late-type stars \cite{Wade2001ASPC248p403}.

However, magnetic fields in hotter OB stars remain difficult to detect. The few photospheric lines present in the optical spectrum and the large intrinsic width of the lines, worsened by the usual fast rotation of those stars, require large-bandwidth and  high signal-to-noise ratio observations to start with, even using LSD.

For those reasons, the magnetic characteristics of the population of neutron star progenitors are currently poorly known, and mostly only by indirect ways, such as non-thermal radio and X-ray emission and variability, peculiar abundances or cyclical variations of photospheric and wind spectral lines (see \citep{Fullerton2003ASPC305p333,Wade2001ASPC248p403} for a review). Nevertheless, those indicators are often interpreted to show that magnetism may be widespread among massive OB stars.

The advent of a new generation of spectropolarimeters such as ESPaDOnS at the Canada-France-Hawaii Telescope (CFHT) and its twin NARVAL at the T\'elescope Bernard-Lyot (TBL) now allows new investigations of magnetic fields in massive OB stars.
ESPaDOnS consists of a polarimetric module located at the Cassegrin focus of the CFHT, linked by optical fibres to the high-resolution echelle spectrometer. A resolution of 65,000 for a spectral range covering from $360\rm\,nm$ to $1\rm\,\mu m$ can be achieved in a single observation. 

A complete circular polarisation observation consists of series of 4 sub-exposures between which the polarimeter quarter-wave plate is rotated back and forth between position angles. This procedure results in exchanging orthogonally polarised beams throughout the entire instrument, which makes it possible to reduce systematic errors.

\subsection{The Orion Nebular Cluster}

\begin{figure}
  \includegraphics[width=0.45\textwidth]{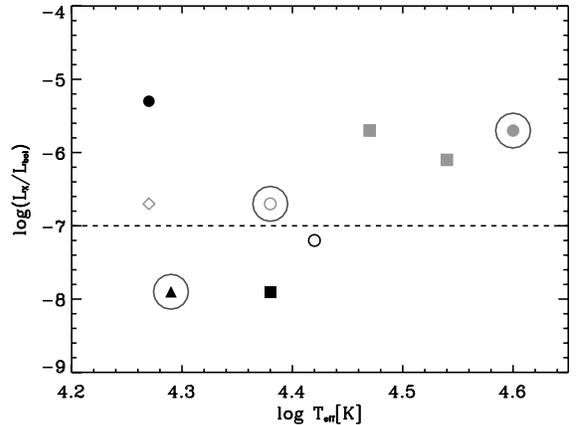}
  \caption{ \label{x} X-ray efficiency as a function of effective temperature. The detected stars are circled. Filled symbols are for stars with indirect indications of the presence of a magnetic field and gray symbols are for confirmed or suspected binaries. Plotting symbols indicate the following properties: circles are for T-Tauri type emission, triangles are chemically peculiar (CP) stars, and the diamond star was not observed. The dotted line indicates the typical efficiency for massive stars. }
\end{figure}

\begin{figure}[t]
  \includegraphics[width=0.95\textwidth]{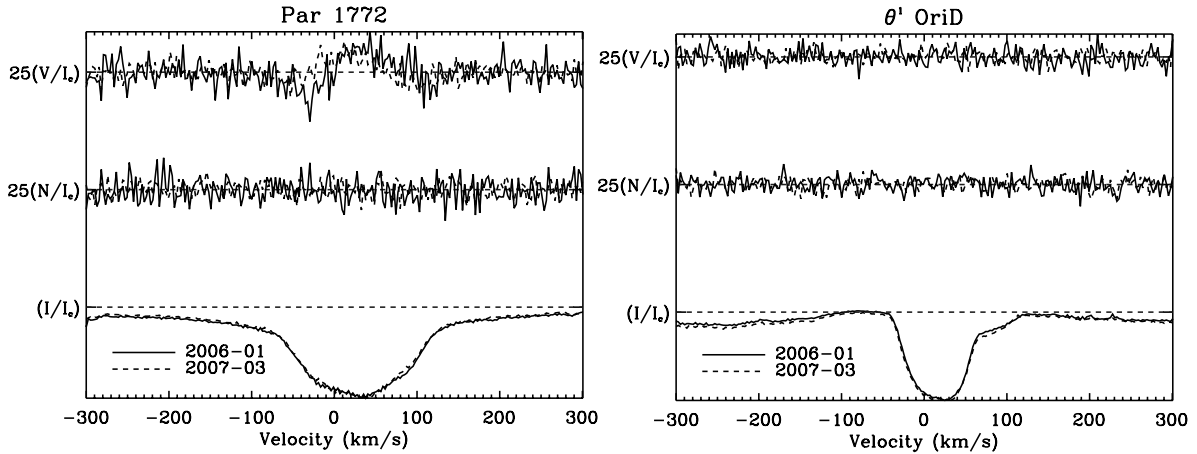}
  \caption{ \label{lsd} Least Squares Deconvolved profiles for Par\,1772 (B2\,V) and $\theta^1\rm\,Ori\,D$ (B0.5\,V). The curves are the mean Stokes I profiles (bottom), the mean Stokes V profiles (top) and the N diagnostic null profiles (middle), in solid line for January 2006 and dashed line for March 2007.}
\end{figure}

\begin{figure}[t]
  \includegraphics[width=0.95\textwidth]{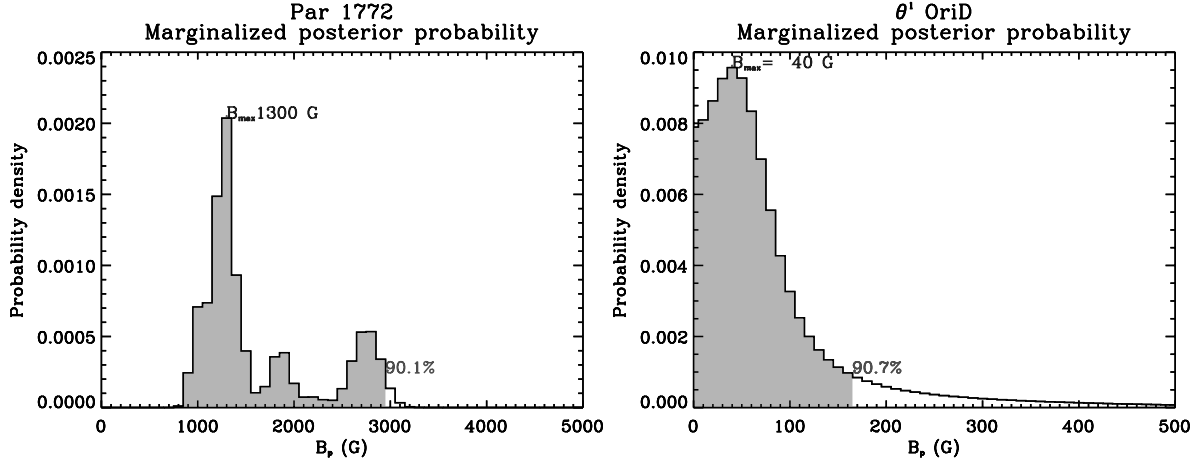}
  \caption{ \label{bayes} Marginalized posterior probability densities for Par\,1772 (one of the 3 detected OB stars) and $\theta^1\rm\,Ori\,D$ (one of the 5 undetected stars). The magnetic field strength 90\% credible region (filled) is [900, 2900]\,G for Par\,1772 and [0, 160]\,G for $\theta^1\rm\,Ori\,D$.}
\end{figure}

The first goal of this project was to explore the connection between magnetic fields and X-ray production in massive OB stars. 
Stellar magnetic fields are well known to produce X-rays in late-type convection stars like the Sun. However, X-ray emission coming from OB stars is often explained by radiative instabilities resulting in a multitude of shocks in their winds \cite{Lucy1980ApJ241p300,Cohen1999ApJ520p833}. 

The {\it Chandra} Orion Ultradeep Project (COUP) was dedicated to observe the Orion Nebula Cluster (ONC) in X-rays. The OBA sample (20 stars) was studied with the goal of disentangling the respective roles of winds and magnetic fields in producing X-rays \cite{Stelzer2005ApJS160p557}. The production of X-rays by radiative shocks should be the dominant mechanism for the subsample of 9 OB stars with strong winds. However, aside from two of those stars, all targets showed X-ray flux intensity and/or variability which were inconsistent with the small shock model predictions (Figure \ref{x}).

For these reasons, we started our investigation with the 9 OB stars of this young star cluster.  They range from B3\,V ($\sim8\rm\,M_\odot$) to O7\,V ($\sim40\rm\,M_\odot$), approximately the mass range from which neutron stars are thought to be formed. 

We conducted spectropolarimetric observations of 8 of those massive stars in January 2006 and March 2007 at CFHT. Using the LSD technique, we found clear Stokes V signatures for 3 stars: the previously-detected $\theta^1\rm\,Ori\,C$, as well as Par\,1772 (shown in Figure \ref{lsd}, along with the non-detection case $\theta^1\rm\,Ori\,D$), and NU\,Ori.

\subsection{Magnetic analysis}

In order to extract the surface field characteristics from the observed Stokes V profiles, we compared them with theoretical profiles for a large grid of dipolar magnetic configurations, calculated with the polarised LTE radiative transfer code Zeeman2 \citep{Landstreet1988ApJ326p967,Wade2001AA374p265}. 

We sampled the 4-dimentional parameter space ($i$, $\beta$, $\phi$, B) which describes a centered dipolar magnetic configuration. In such a model, $i$ is the projected angle of the rotation axis to the line of sight, $\beta$ is the angle between the magnetic axis and the rotation axis, $\phi$ is the rotational phase and B is the polar field strength.

For each configuration, we calculated the reduced $\chi^2$ of the model fit to the observed mean Stokes V profiles. 
Assuming that only the phase may change between two observations of a given star, the goodness-of-fit of a given ($i$, $\beta$, B)-configuration is expressed in terms of Bayesian probability density. This ensures that a good magnetic ($i$, $\beta$, B)-configuration possesses phases that fit both observations, as the rotational period is not known with enough accuracy to determine {\it a priori} the phase difference.  

We can determine the probability density of the field strength, by marginalising over inclination ($i$) and o\-bli\-qui\-ty ($\beta$). 
Then, we extract a 90\% credible region for the field strength of each star (Figure \ref{bayes}) with the technique described by \citet{Gregory2005book}.

Our first results show that any dipolar fields present in the 5 undetected stars are weaker than $\sim200\rm\,G$ with a 90\% confidence. The field strengths of the 3 detected stars are approximately 1-3\,kG.   

\section{Discussion and conclusion}

\begin{figure}
  \includegraphics[width=0.45\textwidth]{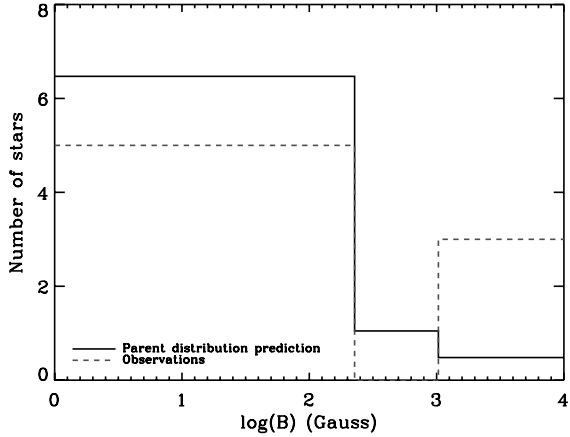}
  \caption{ \label{prob} Predicted magnetic field repartition of 8 randomly-drawn stars according to the predicted distribution of \citet{Ferrario2006MNRAS367p1323} along with the observed distribution of field strengths in the ONC.}
\end{figure}

As an illustrative example, we can compare our new observational results with the predictions made by \citet{Ferrario2006MNRAS367p1323} about the magnetic field distribution of massive stars (8-$45\rm\,M_\odot$) on the main sequence. 

They parametrized the magnetic flux distribution on the main sequence $\chi(\Phi)$ as the sum of two Gaussians, along with the birth spin period of neutron stars. Assuming a complete conservation of magnetic flux, they have calculated the expected properties of isolated radio pulsars.  They used the 1374-MHz Parkes Multi-Beam Survey of isolated radio pulsars in order to constrain the model parameters. They obtained a continuous magnetic field distribution in the progenitor OB stars peaking at $\sim46\rm\,G$ with 5 per cent\footnote{Although the paper states 8\%, recalculation based on the detailed model distribution provided by L. Ferrario gives 5\%.} of the stars having a field in excess of 1\,kG.

Of course, our sample contains only 8 stars, but we can still make some rough comparaisons. 
Taking the predicted field strength distribution, we assume that it is the true parent distribution from which we draw a random sample of 8 stars. We define three possible outcomes: [0-200]\,G, [200-1000]\,G and over 1000\,G, with respective probabilities derived from the parent theoretical distribution (Figure \ref{prob}). According to the multinomial distribution,  the probability of observing the distribution of magnetic field strengths observed in the ONC is below 1\%. 

This result might be interpreted, at first glance, to suggest that the fields of neutron stars are not of fossil origin. However, some points are important to consider: 
\begin{enumerate}
\item The sample of stars may not be representative of a general parent distribution, as the stars all come from the same cluster. This region could be unusually magnetic, especially if the fields of the OB stars themselves are also of fossil origin from the ISM.
\item There may be a fossil component to the magnetic field origin, but the assumed parent distribution is not in fact the true parent distribution because some assumptions are incorrect, or some elements are missing from the model. Examples of such missing physics might be partial flux conservation or the influence of dynamo processes during core collapse.
\end{enumerate}

In order to better explore these possibilities, a larger sample of OB stars, from clusters and from the field, must be studied in order to increase the population of neutron star progenitors with known magnetic properties. Our team has undertaken an extensive spectropolarimetric study of massive stars in other young star clusters to provide these important data.

\begin{theacknowledgments}
  VP acknowledges support from Fonds qu\'eb\'ecois de la recherche sur la nature et les technologies.
  LD acknowledges support from  the Canada Research Chair program and the Discovery Grants programme of the Natural Science and Engineering Research Council of Canada.
  GAW  acknowledges support from the Discovery Grants programme of the Natural Science and Engineering Research Council of Canada.
  
  Based on observations obtained at the Canada-France-Hawaii Telescope (CFHT) which is operated by the National Research Council of Canada, the Institut National des Sciences de l'Univers of the Centre National de la Recherche Scientifique of France, and the University of Hawaii. 
  
  The ESPaDOnS data were reduced using the data reduction software Libre-ESpRIT, written by J.-F. Donati from the Observatoire Midi-Pyr\'en\'ees and made available to observers at the CFHT.
 \end{theacknowledgments}

\bibliographystyle{aipproc}  
\bibliography{onc_pulsar}

\hyphenation{Post-Script Sprin-ger}
\begin{thebibliography}{27}
\expandafter\ifx\csname natexlab\endcsname\relax\def\natexlab#1{#1}\fi
\providecommand{\enquote}[1]{``#1''}
\expandafter\ifx\csname url\endcsname\relax
  \def\url#1{\texttt{#1}}\fi
\expandafter\ifx\csname urlprefix\endcsname\relax\def\urlprefix{URL }\fi
\providecommand{\eprint}[2][]{\url{#2}}

\bibitem[{Alecian} et~al.(2007)]{Alecian2007MNRAS}
E.~{Alecian}, C.~{Catala}, G.~A. {Wade}, J.~F. {Donati}, P.~{Petit}, J.~D.
  {Landstreet}, T.~{Bohm}, J.~C. {Bouret}, S.~{Bagnulo}, C.~{Folsom}, and
  J.~{Silvester}, \emph{\mnras, accepted}  (2007).

\bibitem[{Borra} and {Landstreet}(1980)]{Borra1980ApJS42p421}
E.~F. {Borra}, and J.~D. {Landstreet}, \emph{\apjs} \textbf{42}, 421--445
  (1980).

\bibitem[{Schmidt} et~al.(2003)]{Schmidt2003ApJ595p1101}
G.~D. {Schmidt}, H.~C. {Harris}, J.~{Liebert}, D.~J. {Eisenstein}, S.~F.
  {Anderson}, J.~{Brinkmann}, P.~B. {Hall}, M.~{Harvanek}, S.~{Hawley}, S.~J.
  {Kleinman}, G.~R. {Knapp}, J.~{Krzesinski}, D.~Q. {Lamb}, D.~{Long}, J.~A.
  {Munn}, E.~H. {Neilsen}, P.~R. {Newman}, A.~{Nitta}, D.~J. {Schlegel}, D.~P.
  {Schneider}, N.~M. {Silvestri}, J.~A. {Smith}, S.~A. {Snedden}, P.~{Szkody},
  and D.~{Vanden Berk}, \emph{\apj} \textbf{595}, 1101--1113 (2003).

\bibitem[{Charbonneau} and {MacGregor}(2001)]{Charbonneau2001ApJ559p1094}
P.~{Charbonneau}, and K.~B. {MacGregor}, \emph{\apj} \textbf{559}, 1094--1107
  (2001).

\bibitem[{Wade} et~al.(2007)]{Wade2007MNRAS376p1145}
G.~A. {Wade}, S.~{Bagnulo}, D.~{Drouin}, J.~D. {Landstreet}, and D.~{Monin},
  \emph{\mnras} \textbf{376}, 1145--1161 (2007).

\bibitem[{Wade} et~al.(2005)]{Wade2005AA442pL31}
G.~A. {Wade}, D.~{Drouin}, S.~{Bagnulo}, J.~D. {Landstreet}, E.~{Mason},
  J.~{Silvester}, E.~{Alecian}, T.~{B{\"o}hm}, J.-C. {Bouret}, C.~{Catala}, and
  J.-F. {Donati}, \emph{\aap} \textbf{442}, L31--L34 (2005).

\bibitem[{Catala} et~al.(2007)]{Catala2007AA462p293}
C.~{Catala}, E.~{Alecian}, J.-F. {Donati}, G.~A. {Wade}, J.~D. {Landstreet},
  T.~{B{\"o}hm}, J.-C. {Bouret}, S.~{Bagnulo}, C.~{Folsom}, and J.~{Silvester},
  \emph{\aap} \textbf{462}, 293--301 (2007).

\bibitem[{Alecian} et~al.(2006)]{Alecian2006ArXiv0612186}
E.~{Alecian}, G.~A. {Wade}, C.~{Catala}, S.~{Bagnulo}, T.~{Bohm}, J.~.
  {Bouret}, J.~. {Donati}, C.~P. {Folsom}, J.~D. {Landstreet}, and
  J.~{Silvester}, \emph{ArXiv Astrophysics e-prints}  (2006).

\bibitem[{Folsom} et~al.(2007)]{Folsom2007MNRAS376p361}
C.~P. {Folsom}, G.~A. {Wade}, S.~{Bagnulo}, and J.~D. {Landstreet},
  \emph{\mnras} \textbf{376}, 361--370 (2007).

\bibitem[{Wickramasinghe} and
  {Ferrario}(2005)]{Wickramasinghe2005MNRAS356p1576}
D.~T. {Wickramasinghe}, and L.~{Ferrario}, \emph{\mnras} \textbf{356},
  1576--1582 (2005).

\bibitem[{Gaensler} et~al.(2005)]{Gaensler2005ApJ620pL95}
B.~M. {Gaensler}, N.~M. {McClure-Griffiths}, M.~S. {Oey}, M.~{Haverkorn}, J.~M.
  {Dickey}, and A.~J. {Green}, \emph{\apjl} \textbf{620}, L95--L98 (2005).

\bibitem[{Muno} et~al.(2006)]{Muno2006ApJ636pL41}
M.~P. {Muno}, J.~S. {Clark}, P.~A. {Crowther}, S.~M. {Dougherty}, R.~{de
  Grijs}, C.~{Law}, S.~L.~W. {McMillan}, M.~R. {Morris}, I.~{Negueruela},
  D.~{Pooley}, S.~{Portegies Zwart}, and F.~{Yusef-Zadeh}, \emph{\apjl}
  \textbf{636}, L41--L44 (2006).

\bibitem[{Donati} et~al.(2002)]{Donati2002MNRAS333p55}
J.-F. {Donati}, J.~{Babel}, T.~J. {Harries}, I.~D. {Howarth}, P.~{Petit}, and
  M.~{Semel}, \emph{\mnras} \textbf{333}, 55--70 (2002).

\bibitem[{Donati} et~al.(2006)]{Donati2006MNRAS365pL6}
J.-F. {Donati}, I.~D. {Howarth}, J.-C. {Bouret}, P.~{Petit}, C.~{Catala}, and
  J.~{Landstreet}, \emph{\mnras} \textbf{365}, L6--L10 (2006).

\bibitem[{Heger} et~al.(2005)]{Heger2005ApJ626p350}
A.~{Heger}, S.~E. {Woosley}, and H.~C. {Spruit}, \emph{\apj} \textbf{626},
  350--363 (2005).

\bibitem[{Spruit}(1999)]{Spruit1999AA349p189}
H.~C. {Spruit}, \emph{\aap} \textbf{349}, 189--202 (1999).

\bibitem[{Donati} et~al.(1997)]{Donati1997MNRAS291p658}
J.-F. {Donati}, M.~{Semel}, B.~D. {Carter}, D.~E. {Rees}, and A.~{Collier
  Cameron}, \emph{\mnras} \textbf{291}, 658 (1997).

\bibitem[{Auri\`ere} et~al.(2007)]{Auriere2007}
M.~{Auri\`ere}, G.~A. {Wade}, J.~{Silvester}, F.~{Ligni\`eres}, K.~{Bagnulo},
  K.~{Bale}, B.~{Dintrans}, J.~F. {Donati}, C.~P. {Folsom}, M.~{Gruberbauer},
  A.~{Hui Bon Hoa}, S.~{Jeffers}, N.~{Johnson}, J.~D. {Landstreet},
  A.~{L\`ebre}, T.~{Lueftinger}, S.~{Marsden}, D.~{Mouillet}, S.~{Naseri},
  F.~{Paletou}, P.~{Petit}, J.~{Power}, F.~{Rincon}, S.~{Strasser}, and
  N.~{Toqu\'e}, \emph{\aap, accepted}  (2007).

\bibitem[{Wade}(2001)]{Wade2001ASPC248p403}
G.~A. {Wade}, \enquote{{Zeeman Detection of Magnetic Fields in Hot Stars},} in
  \emph{Magnetic Fields Across the Hertzsprung-Russell Diagram}, edited by
  G.~{Mathys}, S.~K. {Solanki}, and D.~T. {Wickramasinghe}, 2001, vol. 248 of
  \emph{Astronomical Society of the Pacific Conference Series}, p. 403.

\bibitem[{Fullerton}(2003)]{Fullerton2003ASPC305p333}
A.~W. {Fullerton}, \enquote{{Cyclical Wind Variability from O-Type Stars},} in
  \emph{Astronomical Society of the Pacific Conference Series}, edited by L.~A.
  {Balona}, H.~F. {Henrichs}, and R.~{Medupe}, 2003, vol. 305 of
  \emph{Astronomical Society of the Pacific Conference Series}, p. 333.

\bibitem[{Lucy} and {White}(1980)]{Lucy1980ApJ241p300}
L.~B. {Lucy}, and R.~L. {White}, \emph{\apj} \textbf{241}, 300--305 (1980).

\bibitem[{Owocki} and {Cohen}(1999)]{Cohen1999ApJ520p833}
S.~P. {Owocki}, and D.~H. {Cohen}, \emph{\apj} \textbf{520}, 833--840 (1999).

\bibitem[{Stelzer} et~al.(2005)]{Stelzer2005ApJS160p557}
B.~{Stelzer}, E.~{Flaccomio}, T.~{Montmerle}, G.~{Micela}, S.~{Sciortino},
  F.~{Favata}, T.~{Preibisch}, and E.~D. {Feigelson}, \emph{\apjs}
  \textbf{160}, 557--581 (2005).

\bibitem[{Landstreet}(1988)]{Landstreet1988ApJ326p967}
J.~D. {Landstreet}, \emph{\apj} \textbf{326}, 967--987 (1988).

\bibitem[{Wade} et~al.(2001)]{Wade2001AA374p265}
G.~A. {Wade}, S.~{Bagnulo}, O.~{Kochukhov}, J.~D. {Landstreet}, N.~{Piskunov},
  and M.~J. {Stift}, \emph{\aap} \textbf{374}, 265--279 (2001).

\bibitem[{Gregory}(2005)]{Gregory2005book}
P.~C. {Gregory}, \emph{{Bayesian Logical Data Analysis for the Physical
  Sciences: A Comparative Approach with `Mathematica' Support}}, Published by
  Cambridge University Press, Cambridge, UK., 2005.

\bibitem[{Ferrario} and {Wickramasinghe}(2006)]{Ferrario2006MNRAS367p1323}
L.~{Ferrario}, and D.~{Wickramasinghe}, \emph{\mnras} \textbf{367}, 1323--1328
  (2006).

\end{thebibliography}

\end{document}